# Monte-Carlo-based spectral gain analysis for THz quantum cascade lasers

Christian Jirauschek[1,2,a)] and Paolo Lugli[2]

[1]*Emmy Noether Research Group "Modeling of Quantum Cascade Devices", TU München, Arcisstraße 21, D-80333 München, Germany*
[2]*Institute for Nanoelectronics, TU München, Arcisstraße 21, D-80333 München, Germany*
[a)]*Author to whom correspondence should be addressed. Electronic mail: jirauschek@tum.de*

Employing an ensemble Monte Carlo transport simulation, we self-consistently analyze the spectral gain for different THz quantum cascade laser structures, considering bound-to-continuum as well as resonant phonon depopulation designs. In this context, we investigate temperature dependent gain broadening, affecting the temperature performance of THz structures. Furthermore, we discuss the influence of the individual scattering mechanisms, such as electron-electron, impurity and interface roughness scattering. A comparison of the simulation results to experimental data yields good agreement.

## I. INTRODUCTION

The quantum cascade laser (QCL) has become an important source of coherent radiation in the mid-infrared, and its frequency range has been extended to the THz region.[1] However, especially in the THz regime the functionality of QCLs is still quite limited, especially considering the temperature performance. Further improvement requires a thorough understanding of the carrier transport processes in the device, as provided by detailed simulations. Self-consistent Monte Carlo (MC) carrier transport simulations are a well-established tool to analyze and optimize QCL structures.[2-11] Usually such simulations focus on transport properties and level occupations. However, the optical gain achieved in the structure not only depends on the population inversion between upper and lower laser level and the corresponding oscillator strength, but is also affected by the linewidth of the lasing transition. In semiclassical approaches like three-dimensional (3D) MC simulations or rate equation descriptions, the lifetime broadening due to carrier transitions between the different quantum states can be self-consistently evaluated, while collisional broadening due to pure dephasing and cancellation effects due to nondiagonal dephasing contributions can only be considered in fully quantum mechanical approaches.[12-16] On the other hand, the MC approach allows for an inclusion of intercarrier scattering (beyond the mean-field approximation), which has up to now not been achieved for comparable quantum mechanical simulations of QCLs,[12-16] due to the significantly increased computational complexity as compared to single-electron processes like electron-impurity interaction. Electron-electron scattering dominates the intrasubband dynamics and thus greatly influences the kinetic electron distributions within the subbands, and can also strongly affect the intersubband transport, especially for closely spaced energy levels, low temperatures or high doping densities.[17] Besides, due to the relative computational efficiency of the MC approach, also other effects can be fully considered where often approximations have to be used in quantum mechanical simulations, such as neglecting the momentum dependence of scattering mechanisms.[12]

Typically, in 3D MC simulations the spectral gain is evaluated in a phenomenological way, using parameters like the experimentally measured spontaneous emission linewidth.[5,9] This prevents self-consistent gain modeling which can be crucial for analyzing and understanding experimental results. In contrast, in quantum mechanical simulations of QCLs like the nonequilibrium Green's functions (NEGF) or density matrix approach, gain is routinely computed in a self-consistent manner.[12-16] In this context, it is desirable to combine the strengths of fully three-dimensional MC simulations, like the inclusion of electron-electron scattering, with a self-consistent gain analysis. One example is the analysis of the temperature performance of THz structures, where not only the reduced inversion,[11] but also the increased broadening of the gain profile plays a role.[15] Here, MC simulations can bring valuable insights, e.g., by evaluating the role of all important scattering mechanisms, including electron-electron interaction. Also, the relative computational efficiency as compared to NEGF simulations allows us to evaluate a larger parameter field, or to refrain from approximations that are often used in quantum mechanical simulations. For example, the momentum dependence of the scattering matrix elements plays a major role especially for high operating temperatures, an effect often neglected in NEGF simulations to keep the computational effort reasonable.[12]

We self-consistently extract the gain spectra of bound-to-continuum and resonant phonon depopulation THz QCLs from a fully three-dimensional ensemble MC approach, which includes electron-electron scattering. We demonstrate that also within a semiclassical framework meaningful results can be obtained which are in good agreement with experiment. Furthermore, we evaluate the influence of the different broadening mechanisms, specifically discussing the role of carrier-carrier, impurity and interface roughness scattering. We also investigate the temperature dependence of the gain profile. Detailed comparisons to experimental data confirm the validity of our approach.

## II. METHOD

Here, we use a three-dimensional ensemble MC method, accounting for all essential scattering mechanisms. The subband energies and wave functions of the structure are obtained from a Schrödinger-Poisson solver, which is run alternately with the MC program until convergence is reached. Multiple iterations are necessary especially for bound-to-continuum structures, where the charge distribution tends to produce a more significant band bending than for comparable resonant phonon depopulation designs. In the MC solver, inelastic processes due to scattering of electrons with acoustic and longitudinal-optical (LO) phonons as well as elastic impurity and interface roughness scattering are considered on an equal footing. Electron-electron interactions are implemented based on the Born approximation,[18,19] and dynamic screening is also taken into account.[7,8,20] Additionally, nonequilibrium phonon effects are considered, and Pauli's exclusion principle is also accounted for. Periodic boundary conditions are used, i.e., electrons coming out from one side of the device are automatically injected into the equivalent state on the opposite side.[2]

In this MC simulation, the spectral gain is evaluated self-consistently by adding up the Lorentzian gain contributions of all possible optical intersubband transitions from an initial state $|i,\mathbf{k}\rangle$ to a final state $|j,\mathbf{k}\rangle$, thus summing over the corresponding initial and final subbands $i$ and $j$ with energies $E_i$ and $E_j$, and over the in-plane wave vectors $\mathbf{k}$.[14] Each Lorentzian is weighted with the associated optical transition matrix element and the inversion as extracted from the MC analysis for the corresponding $\mathbf{k}$ value.

The Lorentzian lifetime broadening $\gamma_{ij}$ of each transition is extracted self-consistently from the Boltzmann scattering rates. The semiclassical linewidth is given by[21]

$$\gamma_{ij}(\mathbf{k}) = \frac{1}{2}\left[\gamma_i(\mathbf{k}) + \gamma_j(\mathbf{k})\right]. \quad (1)$$

Here, $\gamma_i(\mathbf{k})$ is the total intersubband scattering rate out of a state $|i,\mathbf{k}\rangle$, i.e., we only include scattering between different subbands $i \neq j$. The reason is that for two-dimensional electron systems, the intersubband coherence is ideally not affected by the intrasubband interactions.[22] Thus, the additional consideration of intrasubband scattering events in Eq. (1) would lead to a serious overestimation of the broadening.[23] More generally, nondiagonal correlations between two states lead to a reduction of the corresponding linewidth below the sum of the individual linewidth contributions.[22,23] While such nondiagonal effects can in detail only be considered in fully quantum mechanical simulations, the idealized approach mentioned above, namely assuming full cancellation of broadening associated with intrasubband scattering and neglecting nondiagonal correlations for intersubband events, turns out to work well for very different THz structures, as will be discussed in Section III.

In the following, it is practical to characterize the initial and final states by their kinetic energy $\varepsilon = \hbar^2\mathbf{k}^2/(2m^*)$ instead of $\mathbf{k}$, where $m^*$ is the effective mass. The power gain coefficient $g$ as a function of angular frequency $\omega$ is then given as[24,25]

$$g(\omega) = \frac{e^2\omega}{c\varepsilon_0 n_0 \hbar L} n_E^{2D} \sum_{\substack{i,j \\ E_i > E_j}} |z_{ij}|^2 \int_0^\infty d\varepsilon \left[f_i(\varepsilon) - f_j(\varepsilon)\right] \frac{\gamma_{ij}(\varepsilon)}{\gamma_{ij}^2(\varepsilon) + \left[\omega - (E_i - E_j)/\hbar\right]^2}. \quad (2)$$

For periodic gain structures like QCLs, it is sufficient to consider transitions from states $i$ of a single central period to all available final states (including those in neighboring periods); $L$ is then the length of one period. The constants $e$, $c$ and $\varepsilon_0$ denote the electron charge, vacuum speed of light, and dielectric constant of vacuum, respectively; $n_0$ is the material refractive index, and $n_E^{2D} = m^*/(\pi\hbar^2)$ is the sheet density of electronic states. Furthermore, $z_{ij}$ is the z-operator matrix element between initial and final state, and $f_i(\varepsilon)$ is the occupation probability in a subband $i$ as a function of kinetic energy.

For the numerical evaluation of the spectral gain coefficient, we discretize the integral in Eq. (2) by introducing a kinetic energy grid with energy points $\varepsilon_n$, $n = 1, 2,\ldots$, which divides the energy axis into cells of width $\Delta\varepsilon$ centered around the values $\varepsilon_n$. The number of carriers $N_{i,n}$ in a cell at energy $\varepsilon_n$ in subband $i$ is then related to the corresponding occupation probability $f_i(\varepsilon_n)$ by

$$N_{i,n} = n_E^{2D}\Delta\varepsilon f_i(\varepsilon_n) N/n_s, \quad (3)$$

where $N$ is the total number of simulated electrons in a period, and $n_s$ is the sheet doping density per period. The intersubband outscattering rates $\gamma_i(\varepsilon)$ in Eq. (1) are determined by counting the number of outscattering events $M_{i,n}$ from the respective energy cell to cells in other subbands $j \neq i$ over a given time interval $\Delta t$, yielding

$$\gamma_i(\varepsilon_n) = M_{i,n}/(N_{i,n}\Delta t). \quad (4)$$

## III. RESULTS AND DISCUSSION

In the following, we investigate the material gain spectra of both a resonant phonon depopulation[5,26] and a bound-to-continuum[27,28] THz QCL, using a self-consistent MC-based spectral gain analysis.

### A. Contribution of the various scattering mechanisms

Ensemble MC simulations of THz QCLs routinely include a self-consistent evaluation of electron-phonon and Coulomb interactions as the basic scattering mechanisms. In this context, both electron-electron (e-e) and electron-impurity (e-i) Coulomb interactions play a role and are here considered.[8] As for electron-phonon (e-p) scattering, LO and acoustic phonons are taken into account, with acoustic phonons usually having only a secondary influence. In the following, we show simulation results for the investigated structures and discuss the influence of the different scattering mechanisms.

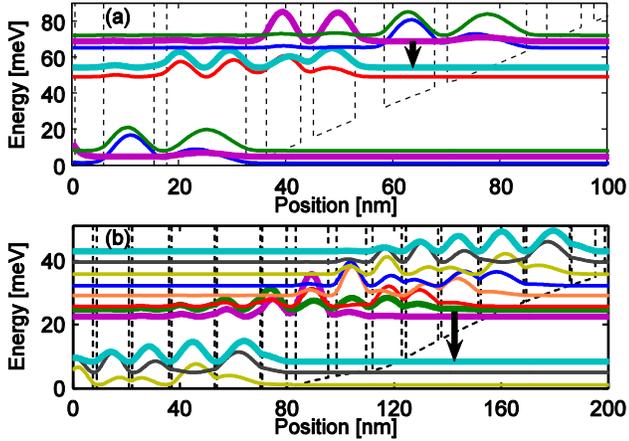

FIG. 1. (Color online) Conduction band profile (dashed lines) and probability densities (solid lines) for two types of QCLs. The laser levels are in bold, and the arrows indicate the lasing transitions. (a) 3.4 THz resonant phonon depopulation structure at 12.25 kV/cm; (b) 3.5 THz bound-to-continuum structure at 3.0 kV/cm.

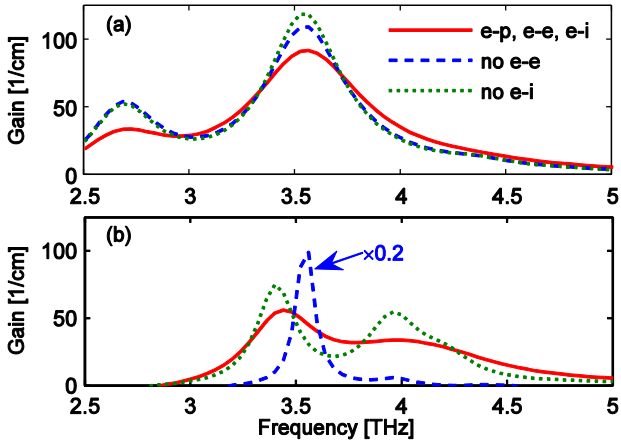

FIG. 2. (Color online) Simulation results for the spectral gain versus frequency, as obtained by evaluating both phonon and Coulomb scattering (solid curves), and neglecting intercarrier (dashed curves) or impurity (dotted curves) scattering. (a) 3.4 THz resonant phonon depopulation structure; (b) 3.5 THz bound-to-continuum structure.

The conduction band structure of the examined THz QCLs is shown in Fig. 1. Figure 2 contains the simulated gain spectra, as obtained when considering all scattering mechanisms discussed above (e-p, e-i, e-e), and when e-e or e-i scattering is neglected. Figure 2(a) contains the results for a 3.4 THz resonant phonon depopulation design at a bias of 12.25 kV/cm,[5,26] and in Fig. 2(b), the gain spectra are displayed for a 3.5 THz bound-to-continuum design[27,28] at a bias of 3.0 kV/cm, where the simulated gain reaches a maximum. For both structures, a lattice temperature of 25 K is assumed. Although they operate at similar frequencies, the two designs are quite different from each other: While the gain can mainly be attributed to a single transition from an upper to a lower laser level for the phonon depopulation structure, two or more transitions significantly contribute to the lasing for the bound-to-continuum design.[27,28] Also the various scattering mechanisms play a different role in each of the structures, as discussed in the following.

For the phonon depopulation structure, the reduced scattering associated with the omission of e-e or e-i interaction in the simulation leads to reduced transition linewidths $\gamma_{if}$, in accordance with Eq. (1), and also to a change in inversion $f_i - f_j$ in Eq. (2). This is reflected by a somewhat sharper gain profile and a slightly increased gain peak, see Fig. 2(a). For the bound-to-continuum design, switching off impurity scattering in the simulation again leads to a reduced transition linewidths $\gamma_{if}$, uncovering two clearly separated peaks in the gain profile (dotted curve in Fig. 2(b)), which correspond to two different lasing transitions. In contrast, the omission of e-e scattering has a much more dramatic effect, yielding an excessively narrowed and raised gain profile (note that the dashed curve in Fig. 2(b) is downscaled by a factor of 0.2). Here, e-e interactions between the closely spaced levels in the minibands clearly play an important role, while in resonant phonon structures, where the energy levels are more separated, the carrier transport is mainly governed by LO phonon scattering. Thus, the inclusion of e-e scattering into the simulation is essential particularly for bound-to-continuum designs. Impurity scattering also plays a role for both structures, but does not have such a profound influence on the gain profiles of the investigated designs.

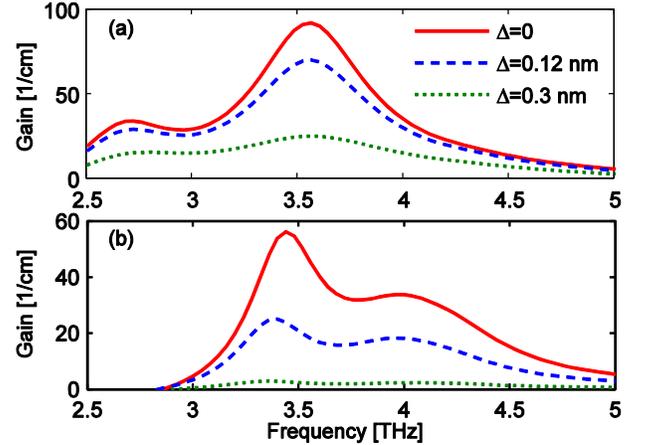

FIG. 3. (Color online) Simulation results for the spectral gain versus frequency, as obtained for different values of the interface roughness mean height $\Delta$. (a) 3.4 THz resonant phonon depopulation structure; (b) 3.5 THz bound-to-continuum structure.

As demonstrated in detailed NEGF simulations, interface roughness can also have a considerable impact on QCL operation.[16] Interface roughness scattering is different from the mechanisms discussed above, in that it cannot be implemented in a completely self-consistent manner, but has to be introduced in a somewhat phenomenological way.[29] The reason is that interface roughness depends crucially on the conditions of growth, thus varying from sample to sample, and its experimental characterization is difficult. Interface roughness is commonly described in terms of a mean height $\Delta$ and a correlation length $\Gamma$.[29] In simulations of GaAs-based THz lasers, often values of around $\Gamma = 10$ nm and $\Delta = 0\dots1$ nm are used.[15,16] In Fig. 3, the gain profile is shown for the same structures and conditions as in Fig. 2, but now including interface roughness scattering in addition to e-p, e-e and e-i interactions. Here we use $\Gamma = 10$ nm,[15] and choose $\Delta = 0$ (no interface roughness scattering), 0.12 nm[15] and 0.3 nm (corresponding to one monolayer of GaAs), respectively. As can be seen in Fig. 3, interface roughness has a similar effect as the impurity scattering discussed above, leading to a further increased broadening and reduced height of the gain profile; the result depends here sensitively on the assumed value for $\Delta$. The phonon depopulation structure in Fig. 3(a) is less sensitive to interface roughness than the bound-to-continuum structure in Fig. 3(b), which we attribute to the fact that the wave functions in the minibands of the latter structure tend to be less localized and thus extend over more interfaces.

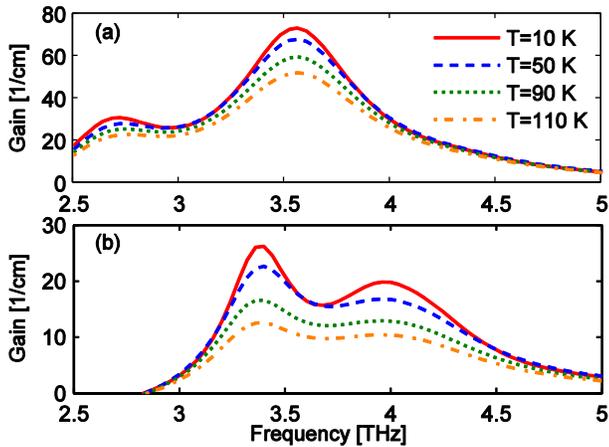

FIG. 4. (Color online) Simulation results for the spectral gain versus frequency, as obtained for different lattice temperatures $T$. (a) 3.4 THz resonant phonon depopulation structure; (b) 3.5 THz bound-to-continuum structure.

## B. Temperature dependence of spectral gain profiles

As discussed in Section I, self-consistent gain profile calculations are especially desirable for investigating the gain dependence on specific parameters, like the operating temperature, where not only the reduced inversion,[11] but also the increased broadening of the gain profile leads to a gain reduction at increasing temperatures.[15] MC-based investigations of the temperature performance have up to now solely focused on the role of the population inversion.[11,30,31] In this context, LO phonon scattering has been identified as the dominant carrier transport channel in resonant phonon structures, largely responsible for the observed reduction in inversion with temperature. The self-consistent gain analysis discussed here makes it possible to assess inversion degradation and gain broadening on an equal footing within the framework of MC simulations.

In the following, we investigate the temperature dependent gain profile for both the phonon depopulation and the bound-to-continuum structure. For the interface roughness scattering, we use typical values of $\Gamma = 10$ nm, $\Delta = 0.12$ nm.[15]

Otherwise, the MC simulation is completely self-consistent. In Fig. 4, the gain profiles are shown for the same structures and conditions as in Figs. 2 and 3, but now for different values of the lattice temperature $T$. For the phonon depopulation structure in Fig. 4(a), the gain broadens from a full width at half-maximum (FWHM) value of 0.72 THz at 10 K to 0.88 THz at 110 K, and the maximum gets reduced from 73 cm$^{-1}$ to 52 cm$^{-1}$. For the bound-to-continuum design in Fig. 4(b), the FWHM increases from 1.09 THz to 1.26 THz, while the peak gets reduced from 26 cm$^{-1}$ to 13 cm$^{-1}$. The temperature broadening of the gain is partly due to the increased LO phonon scattering; furthermore, a reduced screening of the impurity and electron potentials for high temperatures leads to an increased Coulomb scattering.[15] These results clearly show that the MC gain analysis accounts for temperature induced broadening of the gain profile. Moreover, the phonon depopulation structure exhibits a lower temperature sensitivity of the simulated peak gain than the bound-to-continuum structure, in accordance with the experimental observation that phonon depopulation THz designs typically show a superior temperature performance.[32]

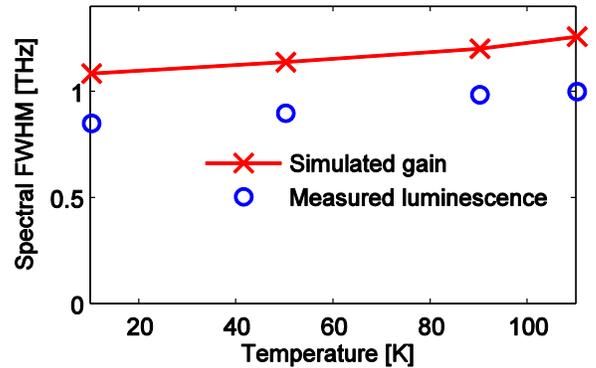

FIG. 5. (Color online) FWHM gain width versus lattice temperature for the 3.5 THz bound-to-continuum structure, as obtained from the simulation and extracted from experimental electroluminescence measurements.

## C. Comparison to experimental data

In the following, we compare the simulated spectral gain profiles to available experimental data, which are largely based on electroluminescence measurements,[26-28] since a direct measurement of the gain profile has only recently been demonstrated and turns out to be quite demanding.[33] In this context, it should however be pointed out that the experimental spontaneous emission linewidth or electroluminescence bandwidth can only serve as a rough estimate of the spectral gain width, since the gain and electroluminescence spectrum are not quite equivalent.[24,27,28]

For the MC analysis, we again assume typical interface roughness values of $\Gamma = 10$ nm and $\Delta = 0.12$ nm.[15] In the following, the simulated temperature broadening of the spectral gain profile is compared to experimental electroluminescence data, which are available for the bound-to-continuum structure.[27,28] In Fig. 5, the computed FWHM values of the gain spectra shown in Fig. 4(b) are plotted versus lattice temperature, together with the corresponding widths extracted from the experimental electroluminescence spectra.[27,28] The theoretically predicted broadening is in excellent agreement with the experimental data, with a simulated ratio of FWHM gain widths at 110 K and 10 K of 1.16, as compared to an experimental value of 1.18. The simulated gain in Fig. 4(b) reaches its maximum at 3.4 THz for all temperatures, which is consistent with the reported lasing at around 3.5 THz.[27,28]

For the phonon depopulation structure, the spontaneous emission linewidth was experimentally determined and can here be used as a benchmark; however, no temperature resolved experimental data are available.[26] The measured value of 0.97 THz agrees well with the theoretical FWHM of 0.72 THz (10 K) to 0.88 THz (110 K), extracted from the spectral gain curves shown in Fig. 4(a). These curves all reach their respective maximum at 3.56 THz, in line with the experimentally observed lasing at around 3.4 THz.[26] The simulated peak gain is 71 cm$^{-1}$ for 25 K, in excellent agreement with a value of 68 cm$^{-1}$ extracted from a previous MC analysis using the spontaneous emission linewidth as experimental input parameter.[5]

The comparison to experimental data shows that the MC-based spectral gain analysis produces meaningful results for quite different THz QCL structures and a broad range of operating conditions. Considering the experimental uncertainties in extracting the gain profile from electroluminescence measurements and the assumptions and approximations used for the simulations, this agreement is all the more remarkable. On the theoretical side, a main error source is the lack of reliable values for the interface roughness parameters, together with the omission of collisional broadening and the simplified treatment of nondiagonal correlations, as discussed in Section II.

## IV. CONCLUSIONS

We have performed an MC-based self-consistent spectral gain analysis of two different THz QCL structures, and have assessed the influence of various scattering mechanisms as well as the temperature broadening. The investigated structures are very different from each other: In the phonon depopulation design, the carrier transport is largely governed by LO phonon scattering, and the gain can be mainly attributed to a single transition. On the other hand, the bound-to-continuum structure is based on minibands consisting of closely spaced energy levels, thus multiple states can contribute to lasing. Also, electron-electron scattering plays a significant role in such minibands, and must properly be accounted for in the simulation. As compared to fully quantum mechanical methods, MC simulations cannot handle collisional broadening, and nondiagonal correlations have to be considered in an idealized way. Nonetheless, a detailed comparison to experimental data shows that the simulation still produces meaningful results for the spectral gain, suitably taking into account all the different effects playing a role in either of the two designs, and adequately reflecting the temperature dependence of the gain profile.

In conclusion, the ensemble MC method has been shown to be suitable for a self-consistent spectral gain analysis of THz QCLs, combining its typical advantages, like its versatility and relative computational efficiency, with extended capabilities beyond mere carrier transport simulations. In particular, due to the inclusion of electron-electron interactions beyond the mean-field approximation, this method is adequate for the detailed investigation of bound-to-continuum structures, where this scattering mechanism plays an important role.

## ACKNOWLEDGMENTS
C.J. acknowledges support from the Emmy Noether program of the German Research Foundation (DFG, JI115/1-1).